\documentclass[12pt]{article}
\usepackage{amssymb}

\begin{document}

\newcommand{\sign}{\mathop{\rm sign}}
\newcommand{\La}{\left\langle}
\newcommand{\Ra}{\right\rangle}

\title{ Self-averaging and On-line Learning
      }

 \author{G. Reents and R. Urbanczik  \\
        Institut f\"ur Theoretische Physik\\
        Universit\"at W\"urzburg \\
        Am Hubland\\
        D-97074 W\"urzburg \\
        Germany
       }
\date{January 1998}
\maketitle

\vspace{-5mm}
\centerline {\bfseries\itshape (to appear in Phys. Rev. Letters)}

\vspace{5mm}
\begin{abstract}
 \setlength{\parindent}{0.0cm}
 Conditions are given under which one may prove that the
 stochastic dynamics of on-line learning can be described by
 the deterministic evolution of a finite set of order parameters
 in the thermodynamic limit. A global constraint on the average magnitude
 of the increments in the stochastic process is necessary to ensure
 self-averaging. In the absence of such a constraint, convergence
 may only be in probability.  
\end{abstract}

On-line learning, introduced in \cite{Ama67,Kin90}, has become an important
paradigm in the analysis of neural networks. Not only has it enabled
the understanding of specific algorithms for a wide range of supervised
learning scenarios and network architectures, 
e.g. \cite{Bie93,Bar95,Saa95}, but one may also derive 
learning algorithms which are highly optimized for a specific problem, 
e.g. \cite{Kin92,Opp96}. Furthermore it is also possible to analyze
unsupervised learning within this framework \cite{Bie94a,Van96}.

The key assumption in on-line learning is that the adaption of the
network is driven, at each time step, by the presentation of a single pattern
which is picked independently of any previous patterns. Thus the
evolution of the state vector of the network is governed by a 
stochastic (Markov) process. However, if the underlying distribution
of patterns is not too complicated, it is possible to characterize
the performance of the network by a few order parameters, and one
expects these parameters to be self-averaging for large networks.
This makes it possible to map the stochastic evolution of the state
vector onto a deterministic evolution of the order parameters, 
thus greatly facilitating a theoretical understanding. 

While the self-averaging properties of the order parameters may
usually be well observed numerically, a rigorous proof of this
crucial fact has so far been lacking.
The goal of this paper is to give conditions on the stochastic dynamics
which ensure that it may be described by deterministic order parameters
in the thermodynamic limit and to clarify the nature of the convergence.
We first review the customary heuristic derivation of the deterministic 
equations in the context of the perceptron learning rule. Next,
a framework for the analysis of on-line learning is established
which is general enough to cover many of the scenarios discussed in the
literature. Within this framework we prove convergence
by exploiting the fact that the thermodynamic limit
is in some ways analogous to a small step size limit in order parameter 
space. (The small step size limit in weight space has been considered in
\cite{Hes93}).
Of course some assumptions about the stochastic process are needed for
the proof,
and the concluding paragraphs discuss examples to show that these
assumptions, while not being necessary for convergence, are nevertheless
quite reasonable. \cite{application}

To fix ideas let us first consider the perceptron learning rule. We are
given a sequence of inputs $\xi^\mu \in {\mathbb R}^N$ and corresponding 
outputs $s^\mu \in \{-1,1\}$ and we assume that the inputs are picked 
independently from some probability distribution. We hope to approximate
the mapping from input to output by a perceptron $s_w$ which
implements the function $s_w(\xi) = \sign(w^T\xi)$ for
$w,\xi \in {\mathbb R}^N$. To this end, we use a new example $(\xi^\mu,s^\mu)$
to update our current estimate of a good weight vector $w^\mu$ by
\begin{equation}
w^{\mu+1} = w^{\mu} + 
    \frac{\eta}{N} \frac{s^\mu-s_w(\xi^\mu)}{2}\xi^\mu\; ,
\label{perc}
\end{equation}
where $\eta\equiv\eta(\mu/N)$ is a possibly time dependent learning rate. 
For simplicity
we assume that the output is itself given by a perceptron with weight vector
$B$, $s^\mu = s_B(\xi^\mu)$. Then the quantity of interest is the
angle between $w^\mu$ and $B$ which may be calculated from the overlaps
$r^\mu = B^T w^\mu$ and $q^\mu = {w^\mu}^T w^\mu$. One 
easily finds recursive equations for $r^{\mu+1}$ and $q^{\mu+1}$ 
using (\ref{perc}). 
These will of course still depend on the entire input sequence
$\{\xi^\mu\}$ but if we assume that 
the input components are picked independently from the normal distribution,
it is straightforward to average over the last input and find:
\begin{eqnarray}
\langle r^{\mu+1} \rangle_{\xi^\mu} &=& 
 r^\mu + \frac{\eta}{N} \frac{1-r^\mu/\sqrt{q^\mu}}{\sqrt{2 \pi}} 
\nonumber\\
\langle q^{\mu+1} \rangle_{\xi^\mu} &=&
 q^\mu + \frac{2\eta}{N} \frac{r^\mu-\sqrt{q^\mu}}{\sqrt{2 \pi}}
       +\frac{\eta^2}{N} \frac{\arccos{(r^\mu/\sqrt{q^\mu})}}{\pi}\;.
\label{semiav}
\end{eqnarray}
Since $r^\mu$ and $q^\mu$ are still stochastic quantities the above 
equations do not seem very helpful. What one would really like to
calculate, is averages over the entire sequence of inputs up to time $\mu$,
for instance:
\begin{equation} 
\langle r^\mu \rangle_\mu = 
\langle r^\mu \rangle_{\xi^0,\xi^1,\ldots,\xi^{\mu-1}}\;.
\end{equation}
At this point it is customary to argue that in the thermodynamic limit,
$N\rightarrow\infty$ but $\mu = {\cal O}(N)$, the overlaps will be 
self-averaging and that  $r^\mu$ will thus be close to 
$\langle r^\mu \rangle_\mu$ for large $N$. Substituting the averages
$(\langle r^\mu \rangle_\mu,\langle q^\mu \rangle_\mu)$ for the
stochastic quantities $(r^\mu,q^\mu)$ in the iteration (\ref{semiav}),
leads to deterministic finite difference equations, and, taking the
large $N$ limit once again, one arrives at the set of differential 
equations:
\begin{eqnarray}
\dot{r} &=& \eta(t) \frac{1-r/\sqrt{q}}{\sqrt{2 \pi}} \nonumber \\
\dot{q} &=& 2\eta(t) \frac{r-\sqrt{q}}{\sqrt{2 \pi}}
       +\eta(t)^2 \frac{\arccos{(r/\sqrt{q})}}{\pi} \;. \label{percdt}
\end{eqnarray} 
One now 
claims that for large $N$ and identical or similar initial conditions 
$(r^{tN},q^{tN})$ will be close to $(r(t),q(t))$.

The main goal of this paper is to make such claims precise and give
conditions under which they are rigorously true. Since we do not want to
confine ourselves to the perceptron, the theory should e.g. cover
learning in multilayer perceptrons as well, we introduce a somewhat more
general framework.  Consider an iteration of the form
\begin{equation}
J^{\mu+1} = J^\mu + N^{-1}f(J^\mu,\xi^\mu) \label{iter}\;,
\end{equation}
where the patterns $\xi^\mu$ are picked independently from a probability
distribution on ${\mathbb R}^{L_N}$. The state vectors $J^\mu$ and the
increments lie in some ${\mathbb R}^{M_N}$. The standard case in on-line
learning is that the input dimension $L_N$ and the system size $M_N$ are
on the order of $N$. Further let
\begin{equation}
{\cal Q}:{\mathbb R}^{M_N} \times  {\mathbb R}^{M_N} \rightarrow V
\subset {\mathbb R}^n
\label{quad}
\end{equation}
be a symmetric, bilinear mapping. The intended interpretation is that
${\cal Q}(J^\mu,J^\mu)$ gives the order parameters of the problem.

It might seem that some order parameters, such as $r^\mu$ in the
above perceptron rule, cannot be obtained by applying a quadratic
form to the state vector. However, by using a larger state vector
this can always be achieved, as well as for instance the transformation
of a nonautonomous system into an autonomous one. In the perceptron case
one may formally augment equation (\ref{perc}) with the following
set of equations:
\begin{equation}
b^{\mu+1} = b^{\mu},\mbox{\ \ \ } \tau_1^{\mu+1} = \tau_1^\mu + 1/N, 
\mbox{\ \ \ }
\tau_2^{\mu+1} = \tau_2^\mu
\label{scratch}
\end{equation}
and fix the initial conditions for the new recursions by 
$
b^0 = B,\mbox{\ } \tau_1^0 = 0,\mbox{\ }\tau_2^0 = 1\,.
$
By aggregating $w,b,\tau_1,\tau_2$ to form a vector $J$
of dimension $2N+2$ the set of equations (\ref{perc},\ref{scratch}) is of
the general form (\ref{iter}). Furthermore we define a bilinear 
symmetric form taking values in ${\mathbb R}^3$ via
$
  {\cal Q}(J,\hat{J}) = (w^T\hat{w},b^T\hat{w},\tau_1\hat{\tau_2})\;,
$
where $\hat{J} = (\hat{w},\hat{b},\hat\tau_1{},\hat{\tau_2})$.
Now ${\cal Q}(J^\mu,J^\mu) = (q^\mu,r^\mu,\mu/N)$ and since we thus obtain
the order parameters of the perceptron rule (\ref{semiav}), this rule
is indeed just a special case of the general framework 
(\ref{iter},\ref{quad}).

Before proceeding with the general theory, 
a word of caution regarding our notation is in order. We are
of course not considering a single stochastic process but a sequence of these.
But to reduce notational overhead we have suppressed the index $N$ in symbols
such as $J^\mu,\xi^\mu,f,{\cal Q}$ and the factor $N^{-1}$ in (\ref{iter}) 
is just an
attempt at suggestive notation. It is, however, crucial that the number $n$
of order parameters (\ref{quad}) and the set of their possible values $V$
be independent of $N$.

Writing ${\cal Q}(J,\hat{J})$ more conveniently as $J*\hat{J}$, the iteration 
(\ref{iter})
yields the following relation for the order parameters 
$Q^\mu = J^\mu * J^\mu\;$:
\begin{eqnarray}
Q^{\mu+1} &=& Q^\mu + N^{-1}F(J^\mu,\xi^\mu)\;, \label{stoc1} \\
F(J,\xi)  &=& 2 J*f(J,\xi) + N^{-1} f(J,\xi)*f(J,\xi)\;.
\label{stoc2}
\end{eqnarray}
For the $Q^\mu$ to be the order parameters of the problem,
the input average of the increment function $F(J,\xi)$ should for large
$N$ converge to a quantity which only depends on $J*J$.
At this point we shall just write
\begin{equation}
\langle F(J,\xi) \rangle_\xi \rightarrow G(J*J)
\end{equation}
and be more precise about the kind of convergence later. That the 
limit $N\rightarrow\infty$ is not just the limit of small step size but a 
thermodynamic one, in which the system size $M_N$ and the input
size $L_N$ may diverge, is important in the definition of $G$: 
The $N^{-1}$ term in the increment function $F$ given by (\ref{stoc2})
will in general give a finite contribution to $G$.  

\newcommand{\muN}{{\lceil tN \rceil}}
We may now associate
to the stochastic process (\ref{stoc1}) the deterministic trajectory
\begin{equation}
\dot{Q} = G(Q) \label{dt}
\end{equation}
and ask whether $Q^{tN}$ converges to $Q(t)$ for large $N$. 
Here and in the sequel, the convention is that a real expression
is rounded when it appears in the position of an integer index, like $tN$
in $Q^{tN}$.
For a given initial condition $Q(0)$, we assume that (\ref{dt}) has a 
solution Q(t) up to certain time $\alpha$. We further require the existence
of a compact set $U$ which contains a neighborhood of the trajectory, 
more formally
\begin{equation}
  \{x\in V: |x-Q(t)| \leq \epsilon \} \subset U\;. \label{eps}
\end{equation}
Note that here and in the sequel we assume $0 \leq t \leq \alpha$.

We are now in the position to state conditions for the convergence of the
stochastic process to the deterministic trajectory:
\begin{description}
\item[(a)] $|G(Q_1) - G(Q_0)| < C|Q_1-Q_0|$ for $Q_1,Q_0 \in U$ and
some constant C.
\item[(b)] 
$|Q^0 - Q(0)| < l(N)$,\\
$|\langle F(J,\xi) \rangle_\xi - G(J*J)| < h(N)$  \ if $J*J \in U$,\\
for suitable functions $l$ and $h$ with 
$\lim_{N\rightarrow\infty}l(N)=\lim_{N\rightarrow\infty}h(N) = 0$.
\item[(c)] $\langle |F(J,\xi)|^2 \rangle_\xi < C^2 (|J*J| + 1)^2$,
for convenience we use the same constant $C$, independent of $N$ as
in condition (a).
\end{description}
The Lipschitz condition (a) makes sure that there is a unique 
deterministic trajectory given the initial value $Q(0)$.
Note that this condition is only required to hold in the
neighborhood $U$ of the deterministic trajectory. Indeed, by considering
e.g. the limit $q\rightarrow 0$ for the perceptron rule (\ref{percdt}),
one sees that even for this simple case no global Lipschitz condition holds.
Condition (b) clarifies the required relationship between
the stochastic process and the deterministic trajectory:
Initial conditions should converge and so should the increments,
at least on average and in the neighborhood $U$. The perhaps most 
interesting condition is (c). In the case of the perceptron learning
rule, the fourth moments of the input distribution must exist, for
the LHS of (c) to be defined. Further the condition implies
\begin{description}
\item[(c')] $|\langle F(J,\xi)\rangle_\xi| < C(|J*J| + 1)$.
\end{description}
and is thus a global constraint on the growth of the increments.

Given these conditions, one may prove convergence by considering 
the difference $\Delta^\mu = Q^\mu - Q(\mu/N)$ between the stochastic and the 
deterministic trajectory. Using the abbreviation 
$g^\mu = N(Q(\mu/N+1/N)-Q(\mu/N))$
the following recursive relation for the variance 
$\sigma^\mu$ of $\Delta^\mu$ is obtained from (\ref{stoc1}):
\begin{eqnarray}
\sigma^{\mu+1} &=& \La|\Delta^{\mu+1}|^2 \Ra_{\mu+1} \nonumber \\
&=& \La  |\Delta^\mu + \frac{1}{N}(F(J^\mu,\xi^\mu) - g^\mu)|^2 \Ra_{\mu+1}
\nonumber \\
&=&  \sigma^\mu + \label{BB} \\
&&2/N \La{\Delta^\mu}^T (F(J^\mu,\xi^\mu) - g^\mu)\Ra_{\mu+1} + \nonumber \\
&&1/N^2\La |F(J^\mu,\xi^\mu)|^2 - 2 F(J^\mu,\xi^\mu)^T g^\mu  +
         |g^\mu|^2 \Ra_{\mu+1} \nonumber
\end{eqnarray}
The next step is to find an upper bound on the increments to $\sigma^\mu$
which depends only on $\sigma^\mu$ itself.
For the $2/N$-term in the above equation
we need to distinguish between $Q^\mu$ being in $U$ or not. So we
rewrite this term as:
\begin{eqnarray}
\lefteqn{\La{\Delta^\mu}^T (F(J^\mu,\xi^\mu) - g^\mu)\Ra_{\mu+1} =}\nonumber\\
 &&
\La (1-\theta(|\Delta^\mu|-\epsilon))
            {\Delta^\mu}^T 
             (\langle F(J^\mu,\xi) \rangle_\xi - g^\mu)
      \Ra_\mu + \nonumber \\
&&
\La   \theta(|\Delta^\mu|-\epsilon)
             {\Delta^\mu}^T 
             (\langle F(J^\mu,\xi) \rangle_\xi - g^\mu)
      \Ra_\mu \label{twoON}
\end{eqnarray}
In the first summand one rewrites the difference as:
\begin{eqnarray}
\langle F(J^\mu,\xi) \rangle_\xi - g^\mu &=&
\langle F(J^\mu,\xi) \rangle_\xi - G(Q^\mu) + \nonumber \\
&& G(Q^\mu) - G(Q(\mu/N)) +\nonumber \\
&& G(Q(\mu/N))-g^\mu\;.
\end{eqnarray}
Expanding the product of ${\Delta^\mu}^T$ with the above RHS,
applying the triangle inequality and then (b) and (a)
one obtains \cite{bounds}
an upper bound: 
\begin{eqnarray}
\lefteqn{\La (1-\theta(|\Delta^\mu|-\epsilon))
            {\Delta^\mu}^T 
             (\langle F(J^\mu,\xi) \rangle_\xi - g^\mu)
      \Ra_\mu \leq} \mbox{\hspace*{3cm}}\nonumber \\
&& (h(N)+C^2/N) \sqrt{\sigma^\mu}  +  C  \sigma^\mu. 
\label{B3}
\end{eqnarray}
To bound the second term in (\ref{twoON}) one uses that the growth
condition (c) implies
\begin{eqnarray}
| {\Delta^\mu}^T  (\langle F(J^\mu,\xi) \rangle_\xi - g^\mu)| 
&\leq& |\Delta^\mu|^2 C +
       |\Delta^\mu| (C^2+C) 
\end{eqnarray}
and thus:
\begin{eqnarray}
\lefteqn{
\La \theta(|\Delta^\mu|-\epsilon)
           {\Delta^\mu}^T 
           (\langle F(J^\mu,\xi) \rangle_\xi - g^\mu)
    \Ra_\mu \leq
}\nonumber \\
&&
 C \sigma^\mu  +
 (C^2+C)
  \La \theta(|\Delta^\mu|-\epsilon)|\Delta^\mu| \Ra_\mu\;.
   \label{B4}
\end{eqnarray}
The remaining average can be further simplified by applying a Tschebyscheff
inequality:
$
\La \theta(|\Delta^\mu|-\epsilon)|\Delta^\mu| \Ra_\mu \leq
\sigma^\mu/\epsilon
$.
Using the growth condition (c) one may bound the $1/N^2$ term
in (\ref{BB}) and combining this with (\ref{B3},\ref{B4}) finally yields:
\begin{eqnarray}
A/N &\geq&
(\sigma^{\mu+1}-\sigma^\mu)/
u\left(\sigma^{\mu}\right)\;, \label{txtb}\\
u(\sigma) &=& (h(N)\sqrt{\sigma} + \sigma)
                + N^{-1}(1+\sqrt{\sigma} + \sigma)\;, \nonumber 
\end{eqnarray}
for a suitable positive $A$ which depends only on $C$ and $\epsilon$.
Note that the bound (\ref{txtb}) holds for any $\mu$ and $N$. 
By rewriting its RHS as an integral and summing over $\mu$
one finds 
$\int_{\sigma^{0}}^{\sigma^{\mu}} u(\sigma)^{-1} {\rm d}\sigma \leq A\mu/N$.
Replacing the term $\sqrt{\sigma}$ in $u(\sigma)$
by its value at the upper limit 
$\sigma^{\mu}$, makes the integral both smaller and simpler and in the end 
yields the following key inequality:
\begin{equation}
\sigma^{\mu} \leq 
  4(N^{-1} + l(N)^2 + h(N)^2)\exp\left(4A\frac{\mu}{N}\right)\;.
\end{equation}
Consequently for $\mu=tN$ the variance decays to zero in the large $N$ limit,
the probability of $Q^{{tN}}$ deviating from the sequence average
$\langle Q^{{tN}} \rangle_{{tN}}$ vanishes and the stochastic process 
is self-averaging. \cite{functions}

Let us next consider relaxing the global constraint (c). Assume a 
situation where (c) holds for $J*J \in U$ but not necessarily outside 
of $U$. We may then replace the update rule $f$ in (\ref{iter}) by
\begin{equation}
\tilde{f}(J,\xi) = \left\{
  \begin{array}{cl}
       f(J,\xi)&\mbox{if\ \ } J*J \in U \\
       0       &\mbox{else.}
 \end{array}
\right.
\end{equation} 
Then, for identical initial conditions, the deterministic trajectory of
$f$ and $\tilde{f}$ will be the same. Further all of the conditions
(a-c) hold for $\tilde{f}$ and this stochastic process is
self-averaging. Since the deterministic trajectory lies strictly in the
interior of $U$ and since the increments $\tilde{f}$ are zero outside of
$U$, this implies that the probability of $\tilde{Q}^{{tN}}$ not lying
in $U$ (for any $t\in [0,\alpha]$) must vanish for large $N$. So given
the same input sequence $\tilde{f}$ will typically give the same result
as the unmodified dynamics $f$, and thus $Q^{{tN}}$ converges to $Q(t)$
in probability. Thus, for this weaker notion of convergence, no global
constraint is needed.

To be able to conclude, however, that in such a situation the stochastic
process given by $f$ is self-averaging, we would have to know that
$Q^{{tN}}$ is well behaved on untypical sequences as well. A simple
example will be sufficient to show that this need not be the case.
Consider the following random walk with a step size which depends on the
length of the current vector:
\begin{equation}
J^{\mu+1} = J^\mu + N^{-1}(Q^\mu-1)\xi^\mu.
\end{equation}
Here we assume $J^\mu,\xi^\mu\in{\mathbb R}^N$, $Q^\mu = |J^\mu|^2$ and the 
components of the $\xi^\mu$ are picked independently from the normal 
distribution. Averaging the self-overlap $Q^{\mu+1}$ with respect to the 
last input yields 
\begin{equation}
\langle Q^{\mu+1} \rangle_{\xi^\mu} =
  Q^\mu + N^{-1}(Q^\mu-1)^2 \label{Qsp}
\end{equation}
and condition (c) is violated.
The deterministic trajectory is given by $\dot{Q} = (Q-1)^2$ and while
it is defined for all times if $Q(0)\leq 1$, it will diverge at some finite
time if the initial condition has $Q(0) > 1$.  Consequently one will
expect $Q^{tN}$ to diverge with $N$ if $Q^0$ is greater than $1$ and $t$ is 
sufficiently large. To obtain a lower bound on this divergence, first note
that by convexity of the RHS in (\ref{Qsp})
\begin{equation}
\langle Q^{\mu+1} \rangle_{\mu+1} \geq 
\langle Q^\mu\rangle_\mu   + N^{-1}(\langle Q^\mu\rangle_\mu -1)^2\;.
\end{equation}
Setting $\tilde{Q}^\mu = \langle Q^\mu\rangle_\mu-1$ yields
$\tilde{Q}^{\mu+1} \geq \tilde{Q}^\mu + N^{-1}(\tilde{Q}^\mu)^2$ 
and dropping the first summand allows us to solve the recurrence and find
\begin{equation}
\tilde{Q}^\mu \geq (\tilde{Q}^{\mu_0}/\sqrt{N})^{\textstyle 2^{\mu-\mu_0}}\;.
\label{grow}
\end{equation}
Thus $\tilde{Q}^\mu$ (and $\langle Q^\mu\rangle_\mu$) will increase 
super-exponentially with $\mu$ if ever $\tilde{Q}^{\mu_0}$ becomes
larger than $\sqrt{N}$.

Let us now consider the dynamics for an initial condition
$Q^0 = Q(0) = 0$. By the general theory presented above $Q^{tN}$ will 
with increasing $N$ converge in probability to $Q(tN)$ for any fixed $t$.
There is, however, a small probability of making a large first step.
In particular, the probability of having  $Q^1 > N$ is
larger than $\exp(-P(N))$, where $P(N)$ is a suitable polynomial in $N$.
Whenever such a rare event ($Q^1 > N$) occurs, due to (\ref{grow})
the following steps lead to an extremely fast growth. Consequently
$\langle Q^{tN} \rangle_{tN}$ diverges with $N$ for any positive $t$
and the stochastic dynamics is not self-averaging in the thermodynamic
limit.

While we believe that the conditions imposed on the stochastic process
are not overly restrictive, they are not necessary for self-averaging to
hold. A case in point is the perceptron rule (\ref{perc}). While conditions
(a-c) are true for an initial condition with $q(0) > 0$, the Lipschitz
condition (a) is violated for $q(0)=0$ and it is not possible to define
$r/\sqrt{q}$ in a manner that would make the RHS of (\ref{percdt}) continuous
in the point $q(0)=0$. Nevertheless, numerical simulations indicate
that the perceptron rule is self-averaging for this initial condition.
But this property is highly dependent on fine details of the input
distribution: Instead of always choosing Gaussian inputs, consider
presenting a Gaussian input only in one half of the steps and else presenting
some fixed vector. More formally, let the input $\xi$ be the random variable
$\xi = \chi\tilde{\xi} + (1-\chi)N^{-1}b$, where $\tilde{\xi}$ has normally
distributed components, $\chi$ is $0$ or $1$ with equal probability and
$b$ is a fixed vector with $|b|\leq 1$. The deterministic trajectory
does not depend on the specific choice of $b$ and is, up to a factor of
$1/2$, still given by (\ref{percdt}). If we choose $b=0$, the stochastic
process is essentially the same as for plain Gaussian inputs, except that,
on average, the weight vector does not change in every second step. But now
consider the choice $b=B$ and assume that $\sign(0)= 0$. For $q^0 = 0$
in one half of the cases perfect generalization will be achieved in the first
step and subsequently the weight vector will not change. However,
if the initial input is Gaussian ($\chi^0 = 1$), any subsequent presentation
of $B$ as input will not change the weight vector since
we already have positive overlap with the teacher and the subsequent
dynamics will be the same as for the choice $b=0$. So, for $b=B$, the first
step is crucial and the on-line dynamics is not self-averaging.  

While the above example is rather construed, it does nevertheless show
that the self-averaging properties of on-line learning can depend on rather
minute details of the stochastic process if the Lipschitz condition 
is violated. Consequently we believe that it is difficult to find easily
verifiable conditions for self-averaging which are much weaker than the
ones presented here.  

The authors wish to acknowledge helpful discussions with M. Biehl,
W. Kinzel, and M. Opper. The work of one of us (R.U.) was supported by the 
{\em Deutsche Forschungsgemeinschaft}.


\begin{thebibliography}{99}
\bibitem{application} 
Indeed, our convergence conditions apply to most of the on-line 
scenarios discussed in the literature. To be specific, they hold 
e.\,g. for backpropagation in MLP's as discussed in \cite{Saa95}, 
assuming Gaussian inputs and exluding the pathological case of a 
vanishing initial weight vector.
\bibitem{bounds}
Here and in the sequel we also assume that $C$ is large enough for
the following conditions to hold: 
$|g^\mu - G(Q(\mu/N))| \leq C^2/N \mbox{,\ \ } |g^\mu| \leq C
\mbox{ and \ \ }|Q(\mu/N)|+1 < C$. That such a $C$ can always be found,
is a consequence of standard results in the theory of differential equations
and the fact that $G$ is Lipschitz on the compact set $U$.
\bibitem{functions} 
Further, any quantity which is a function of the $Q^{tN}$ 
(e.g. an angle between vectors) will be self-averaging as well, 
if the function is Lipschitz. In case the function is only 
continuous, convergence in probability is still assured.
\end{thebibliography}
\end{document}